\documentclass[journal]{IEEEtran}

\usepackage{xcolor,soul,framed} 

\colorlet{shadecolor}{yellow}
\usepackage[pdftex]{graphicx}
\graphicspath{{../pdf/}{../jpeg/}}
\DeclareGraphicsExtensions{.pdf,.jpeg,.png}
\usepackage{amsfonts}
\usepackage{multirow}

\usepackage[utf8x]{inputenc}

\usepackage[cmex10]{amsmath}
\usepackage{amssymb}
\usepackage{array}
\usepackage{mdwmath}
\usepackage{mdwtab}
\usepackage{eqparbox}
\usepackage{url,balance,comment}
\usepackage[noadjust]{cite}
\usepackage{tikz}
\usepackage{tikzscale}
\usepackage{graphicx}
\usepackage{tkz-euclide} 
\usetikzlibrary{shapes,arrows} 
\usepackage{tikz-dimline}
\usetikzlibrary{arrows.meta}
\usetikzlibrary{external}
\ifCLASSOPTIONcompsoc
\usepackage[caption=false,font=normalsize,labelfon
t=sf,textfont=sf,subrefformat=parens,labelformat=parens]{subfig}
\else
\usepackage[caption=false,font=footnotesize,subrefformat=parens,labelformat=parens]{subfig}
\fi
\usepackage{url}

\usetikzlibrary{arrows}
\usetikzlibrary{automata,positioning}
\usetikzlibrary{patterns}
\usepackage[ruled,vlined,linesnumbered]{algorithm2e}

\usepackage{color}

\usepackage{mathtools}
\usepackage[super]{nth}
\usepackage{tabstackengine}
\usepackage{amsmath}
\usepackage{setspace}

\usepackage[noadjust]{cite}
\SetKwInput{KwData}{Input}
\SetKwInput{KwResult}{Output}

\newcommand{\norm}[1]{\left\lVert #1 \right\rVert}

\usepackage{pgfplots}
\pgfplotsset{compat = newest}

\def\crosssize{0.065in}
\tikzset{cross/.style={cross out, draw, minimum size=\crosssize,thick, inner sep=0pt, outer sep=0pt}}

\graphicspath{{figures/}}

\begin{document}
\bstctlcite{IEEEexample:BSTcontrol}
    \title{High-Cardinality Hybrid Shaping for 4D Modulation Formats in Optical Communications Optimized via End-to-End Learning}
  \author{Vinícius~Oliari,~\IEEEmembership{Student Member,~IEEE,}
      Boris~Karanov,~\IEEEmembership{Member,~IEEE},
      Sebastiaan~Goossens,~\IEEEmembership{Student Member,~IEEE},
      Gabriele~Liga,~\IEEEmembership{Member,~IEEE},
      Olga Vassilieva, Inwoong Kim, Paparao Palacharla, Chigo Okonkwo, Alex Alvarado,~\IEEEmembership{Senior Member,~IEEE}
  \thanks{
    V. Oliari, B. Karanov, G. Liga, S. Goossens, C. Okonkwo and A. Alvarado are with the Signal Processing Systems (SPS) Group, Department of Electrical Engineering, Eindhoven University of Technology, 5600 MB Eindhoven, The Netherlands (e-mail: v.oliari.couto.dias@tue.nl).
  
  Olga Vassilieva, Inwoong Kim and Paparao Palacharla are with Fujitsu Network Communications, Inc., Richardson, 75082 TX, USA.
  
    This work is supported by the Netherlands Organisation for Scientific Research (NWO) via the VIDI Grant ICONIC (project number 15685). The work of A. Alvarado and S. Goossens has received funding from the European Research Council (ERC) under the European Union's Horizon 2020 research and innovation programme (grant agreement No 757791). The work of G.~Liga is supported by the EuroTechPostdoc programme under the European Union's Horizon 2020 research and innovation programme (Marie Sk\l{}odowska-Curie grant agreement No 754462).}
}  

\markboth{Preprint, \today}{Oliari, V., \textit{et al.}: ML and Shaping}

\maketitle

\begin{abstract}
In this paper we carry out a joint optimization of probabilistic (PS) and geometric shaping (GS) for four-dimensional (4D) modulation formats in long-haul coherent wavelength division multiplexed (WDM) optical fiber communications using an auto-encoder framework. We propose a 4D 10 bits/symbol constellation which we obtained via end-to-end deep learning over the split-step Fourier model of the fiber channel. The constellation achieved 13.6\% reach increase at a data rate of approximately 400 Gbits/second in comparison to the ubiquitously employed polarization multiplexed 32-QAM format at a forward error correction overhead of 20\%.

\end{abstract}


\IEEEpeerreviewmaketitle

\section{Introduction}

\IEEEPARstart{J}{oint} optimization of transmitter and receiver digital signal processing (DSP) blocks, such as coding, modulation, and equalization, can be challenging in optical communications. One possible way of jointly optimizing these blocks is using an autoencoder structure, as first proposed for wireless communications in \cite{Jakob2017}. In the context of optical fiber communications, autoencoders were first introduced and experimentally demonstrated for highly-nonlinear dispersive short-reach links \cite{Boris2018AE}. The concept was then also applied for long-haul coherent fiber systems, optimizing the modulation format using a simple approximated model of the nonlinear channel for 2D constellations \cite{Kadir2020,Vladislav2021,jones2019endtoend}. Deep learning was also used for obtaining 4D modulation formats in \cite{Essiambre2020}, where optimization was performed in the linear regime and the performance was tested for the nonlinear optical fiber channel for $7$ bits/symbol and constant modulus constellations. Beyond deep learning, many strategies for geometric shaping (GS) optimization have been proposed in the literature. In \cite{Bin2021}, 4D $7$ bits/symbol modulation formats were also considered.

In addition to optimizing the modulation format, another technique for improving data rates is to optimize the symbol probabilities. In \cite{Sillekens2018}, symbol probabilities for polarization-multiplexed (PM) quadrature amplitude modulation (QAM) constellations were optimized on a simplified model and then tested on the nonlinear fiber channel. Finding the optimal modulation format geometry and symbol probabilities is a challenging task and an open problem in optical fiber communications \cite{dar2014}. On the other hand, as shown in \cite{StarkJointPSGS,aref2021endtoend} for the wireless channels, deep learning and artificial neural networks (ANN) can provide viable alternatives to the conventional approaches and enable joint GS and probabilistic shaping (PS), which is called hybrid GS and PS.


In this paper, we perform hybrid GS and PS for optical fiber communications using the split-step Fourier method (SSFM) for dual polarization as our channel model. SSFM is one of the most accurate representations of the nonlinear propagation in optical fibers and has already been used in the autoencoder framework \cite{Uhlemann2020}. However, in \cite{Uhlemann2020}, the waveform is optimized together with a 2D constellation format. The optimized constellations in this paper are in 4D and have no constraints in terms of energy or symmetries. This is the first time, to the best of our knowledge, that a 4D 10 bits/symbol constellation has been optimized for the nonlinear optical channel using SSFM data. The results are validated via generalized mutual information (GMI) \cite{Alvarado2018AIR}. 
The optimized constellation demonstrates gains in terms of GMI and transmission distance with respect to both PM-32QAM and PM-64QAM, which are 10 and 12 bits/symbol constellations, respectively.

\section{End-to-end System and Performance Metric}

\begin{figure*}[t!]
    \centering
    \input{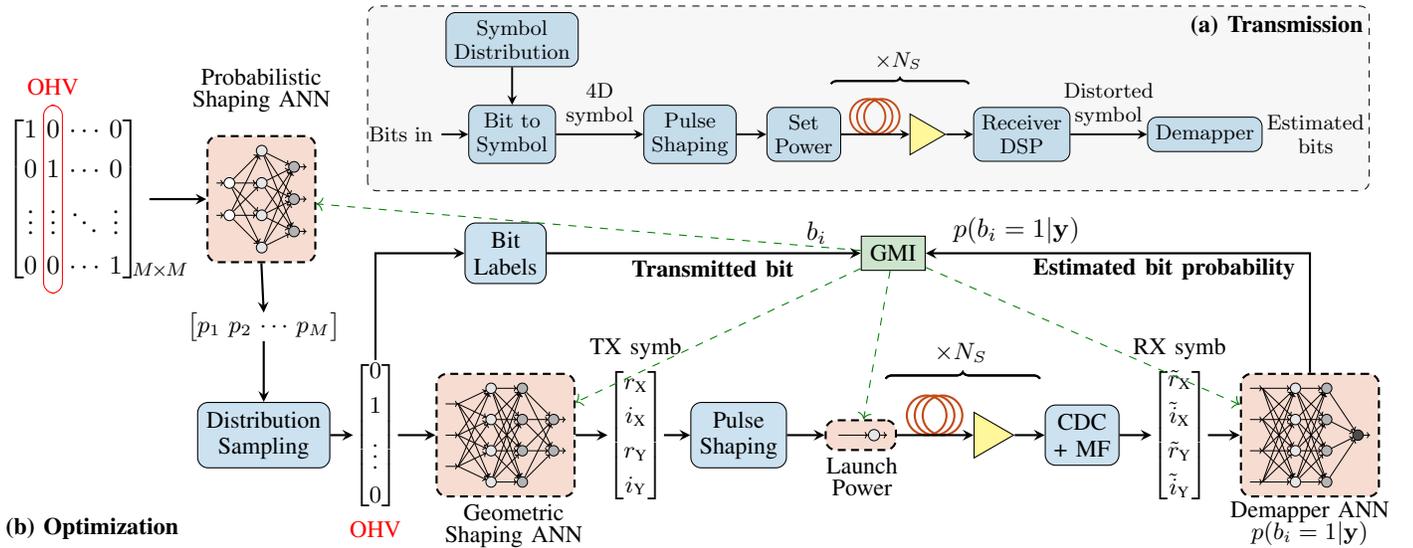}
    \caption{(a) Transmission system used to validated the performance of different modulation formats. (b) Autoencoder setup used to obtain optimal constellation and symbol probabilities. The blocks for probabilistic shaping, geometric shaping, and demapper are built using dense neural networks.}
    \label{fig:system}
\end{figure*}


The GS and PS optical fiber system considered in this paper is shown in Fig.~\ref{fig:system} (a), where the input bits are associated with a transmitted symbols drawn from a certain symbol distribution. These symbols can use a different GS scheme for each transmission case. For example, regular QAM formats or deep learning optimized constellations can be used in the transmission. These symbols are upsampled at 16 samples per symbol and pulse-shaped by a root-raised cosine filter with 0.01 roll-off. The resulting filtered waveform is scaled to achieve a given launch power and propagated in the fiber. The fiber is modeled by the Manakov equation for dual polarization \cite{WaiManakovEq}
\begin{equation}
    \dfrac{\partial \mathbf{A}(t,z)}{\partial z} = -\dfrac{j\beta_2}{2} \dfrac{\partial^2 \mathbf{A}(t,z)}{\partial t^2} +j\dfrac{8}{9}\gamma e^{-\alpha z} \norm{\mathbf{A}(t,z)}^2 \mathbf{A}(t,z),
\end{equation}
where $\mathbf{A}(t,z)$ is the dual-polarization transmitted signal, $\beta_2$ is the group-velocity dispersion parameter, $\gamma$ is the nonlinear Kerr coefficient and $\alpha$ the attenuation coefficient. For the simulations carried on this paper, we transmitted 5 channels, each of them at $50$ Gbaud and channel spacing of $51.5$ GHz. We have chosen a small difference between symbol rate and channel spacing in order to obtain high spectral efficiency. We considered $\beta_2 = -21.67$ ps$^2$/km, $\gamma=1.2$ 1/W/km, and $\alpha= 0.2$ dB/km. The transmission is performed over $N_{S}$ spans of $80$ km each, where an EDFA with noise figure of $5$~dB is used at the end of each span. After propagation, the receiver applies chromatic dispersion compensation, matched filter, and sampling to the received signal, resulting in the received symbols. The received symbols are given to a demapper, which estimates the transmitted bits. In the system of Fig.~\ref{fig:system}~(a), we compare the different modulation formats in terms of the generalized mutual information (GMI) \cite{Alvarado2018AIR}.
The equation for the GMI was obtained by adapting \cite[Eq. (36)]{Alvarado2018AIR} for different a priori probabilities and 4D formats.

\begin{figure*}
    \centering
    \includegraphics[width=1\textwidth]{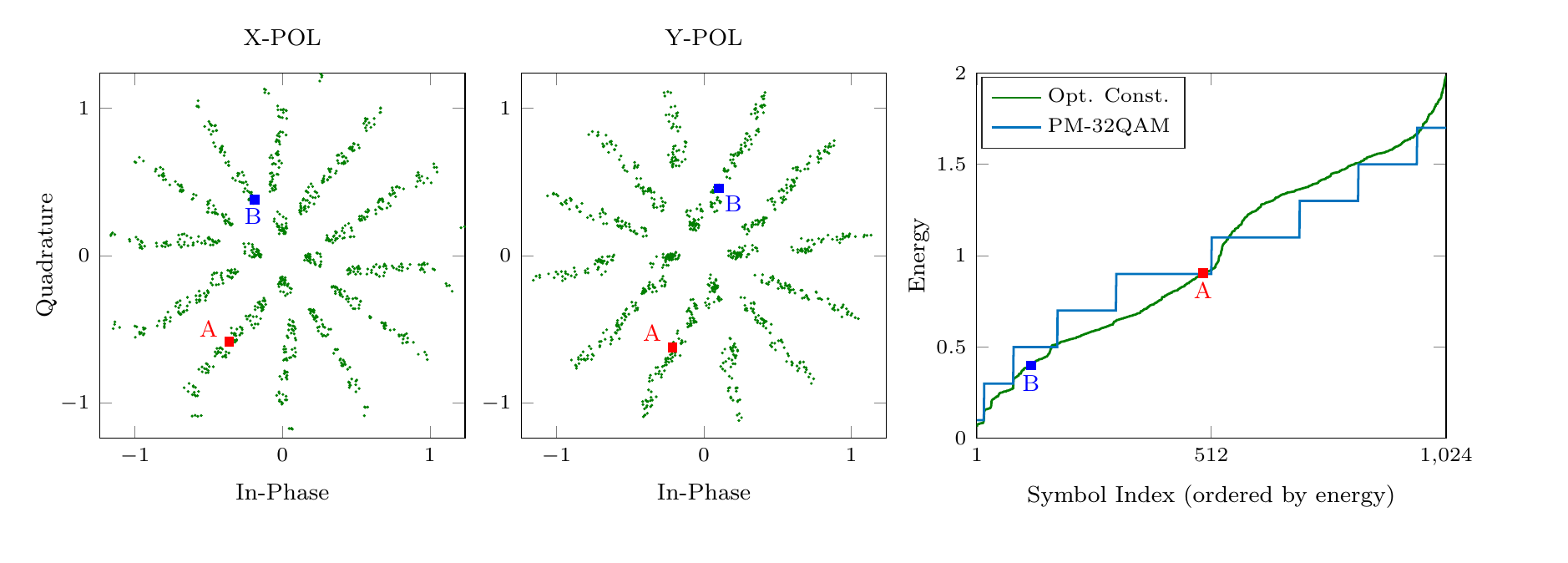}
    \caption{Optimized 4D constellation via the autoencoder structure of Fig.~\ref{fig:system} (a) for X and Y polarizations at optimum launch power and 4000 km (left). Energy per symbol index (right) for the optimized constellation and PM-32QAM. }
    \label{fig:Const}
\end{figure*}

Fig.~\ref{fig:system} (b) shows the autoencoder system used to optimize the constellation geometry and symbol probabilities. The joint optimization of PS and GS follows the same idea described in \cite{StarkJointPSGS}. The optimization system has the same fiber link as the transmission system described for Fig.~\ref{fig:system} (a). The symbol probabilities are obtained from an ANN, whose input are one-hot vectors (OHVs) representing each possible symbol. The ANN output is the logits of the respective symbol probability $[p_1,p_2,\cdots,p_M]$ and each of the five channels has its respective ANN. After obtaining all symbol probabilities, a distribution sampling based on the straight-through Gumbel-Softmax estimator \cite{GumbelSoftmax2017} outputs OHVs sampled according to the previously obtained symbol probabilities. These OHVs are the input of another ANN, labeled as Geometric Shaping in Fig.~\ref{fig:system} (b). For each possible OHV, this ANN outputs a transmitted 4D symbol represented by $[r_{\text{X}}, i_{\text{X}}, r_{\text{Y}}, i_{\text{Y}}]$, where $r_{\text{X}}$ and $i_{\text{X}}$ are the in-phase and quadrature components of the X polarization, and $r_{\text{Y}}$ and $i_{\text{Y}}$ are the respective components of the Y polarization. Each of the five channels has its respective ANN for the 4D symbol mapping. The transmitted symbols are pulse-shaped by the same filter as in Fig.~\ref{fig:system} (a) and the transmitted power for each channel is individually learned. For the training procedure, we used the SSFM with a fixed number of spans $N_{S}=50$. The received signal is processed by the DSP, resulting in the received symbols $\mathbf{y} = [\tilde{r}_{\text{X}}, \tilde{i}_{\text{X}}, \tilde{r}_{\text{Y}}, \tilde{i}_{\text{Y}}]$. These symbols are the input of a set of ANNs, labeled as Demapper, where each of them estimates the bit probabilities $p(b_i=1|\mathbf{y})$ for a specific bit $i$. Each channel has its own set of Demapper ANNs. This bit probability estimation is similar to the procedure described in \cite{jones2019endtoend}, in which $p(b_i=1|\mathbf{y})$ is used to estimate the GMI via \cite[Eq.~(2)]{jones2019endtoend}
\begin{align}\label{eq:gmi}
    \text{GMI} &\approx {H}(\mathbf{X}) + \dfrac{1}{K} \sum_{k=1}^K \sum_{i = 1}^m h_b(b_{i,k},r_{i,k}), \end{align}
    where ${H}(\mathbf{X})$ is the entropy of the random vector $\mathbf{X}$ of the transmitted symbols, $b_{i,k}$ is the $i$-th bit of the $k$-th transmitted symbol, $r_{i,k}$ is the estimated probability $p(b_{i,k} = 1 | \mathbf{y}_k)$ given the received symbol $\mathbf{y}_k$, $m$ is the number of bits per symbol, and $K$ is the number of transmitted symbols. In \eqref{eq:gmi}, we use the function $h_b$ given by
    \begin{align}
    h_b(b_{i,k},r_{i,k}) &= b_{i,k} \log(r_{i,k}) + (1-b_{i,k}) \log(1-r_{i,k}).
\end{align}
 The GMI estimation in \eqref{eq:gmi} is different from the one used in \cite[Eq. (36)]{Alvarado2018AIR}, for example. Computing the expression in \cite[Eq. (36)]{Alvarado2018AIR} for 4D modulation formats with 10 bits/symbol is more computational demanding than \eqref{eq:gmi}, which is why we used \eqref{eq:gmi} for the optimization process. On the other hand, since \eqref{eq:gmi} demands an ANN to compute the bit probabilities, the adaptation of \cite[Eq. (36)]{Alvarado2018AIR} for different a priori probabilities is used to compare the performance of the different modulation formats.


In our system, the geometric shaping ANN has 3 layers with ReLU activation and a final layer with linear activation, all densely connected and with 256 nodes. The probabilistic shaping ANN has a similar structure, but with 2 layers with ReLU activation instead of 3, and also an output layer with linear activation function. We use a linear activation function for the output layer since first we obtain the logits and then the symbol probabilities. The $m$ demapper ANNs are composed by 4 densely connected layers with 256 nodes each, where the first three activation functions are ReLU and the last one is a sigmoid, since their outputs are bit probabilities. We sum the GMI estimation of each channel and use the result as the loss function for the system. The training was performed using an ADAM optimizer as in \cite{Boris2018AE} with learning rate 0.0005, and exponential decays for first- and second-order moments given by 0.9 and 0.999, respectively. The batch size was 2 and the number of symbols for each channel per mini-batch was $2^{13}$. The simulations were run in single precision for approximately 300000 optimization iterations. The low batch size, number of symbols per channel, and the choice of single precision was due to memory limitations in the simulation.

We used two different systems, one for transmission and one for optimization. This is to emphasize that the ANNs present in the optimization are only used to obtain the symbol probabilities and constellation points. Therefore, the system complexity is the same as a standard 4D transmission\begin{footnote}{The demapper, which could be a burden for 4D modulation formats, could be implemented with a computationally efficient ANN \cite{Essiambre2020}. However, the tested transmission system was simulated without an ANN.}\end{footnote}.

\section{Results}

Fig.~\ref{fig:Const} shows the normalized constellations in X and Y polarizations obtained via the autoencoder structure of Fig.~\ref{fig:system} (b). The illustrated constellation is respective to one of the outer channels. Although these constellations suggest symmetries in each polarization, these patterns were all learned by the system without any constraint applied. Interestingly, the probabilities for the learned constellations of all channels converged to uniform probabilities. This uniform distribution might indicate that the geometry of the constellation is more important than its probability distribution at optimum launch power for this relatively large number of constellation points in 4D. Fig.~\ref{fig:Const} also depicts the energy of each symbol for the learned constellation and for PM-32QAM, which also has 10 bits/4D-symbol. As shown in Fig.~\ref{fig:Const}, the symbols have a wide energy variation, indicating that an energy constrain might not be necessary when optimizing for the nonlinear optical channel. The learned constellation is not polarization-multiplexed as in the case of PM-32QAM. The energy of two specific symbols (A and B) is also highlighted.

Fig.~\ref{fig:gmi} depicts the average GMI over the 5 channels for the optimized constellation and traditional QAM constellations. For these results, all channels were propagated with same launch power since using the different learned launch powers per channel by the autoencoder did not provide additional gains. The left part of Fig.~\ref{fig:gmi} shows the GMI versus distance at 4000 km ($N_S=50$ spans). At this distance, the proposed constellation achieves an average of approximately 400 Gbits/symbol or 8 bits/symbol, which yields a spectral efficiency of approximately 7.76 bits/s/Hz. The proposed constellation also outperforms PM-32QAM by 0.3 bits/symbol at optimum launch power. In addition, this constellation shows better performance than PM-32QAM in the linear regime. The performance of the optimized constellation is close to the performance of PM-32QAM for $N_S=44$ spans (3520~km), which indicates a $13.6\%$ reach increase.

\begin{figure*}
    \centering
    \scalebox{1}{\input{figures/GMIvsLP.tikz}}
    \caption{GMI results versus input power (left) and net rates versus distance (right). The learned constellation is compared with standard QAM modulation formats in a multiple-span fiber link. In the right figure, the transmission distance is varied by changing the number of spans.}
    \label{fig:gmi}
\end{figure*}

Fig.~\ref{fig:gmi} (right) shows the net rate per channel versus distance for the optimized constellation, PM-32QAM, PM-64QAM, and PM-PS64QAM. The latter modulation format corresponds to probabilistic-shaped PM-64QAM using a Maxwell–Boltzmann distribution. These results show that the optimized constellation outperforms both PM-16QAM and PM-32QAM for all the displayed distances, and outperforms PM-64QAM for distances higher than 3500 km. Between 4000 and 4800 km, the forward error correction (FEC) overhead (OH) to obtain the data rate given by the optimized constellation is between 20\% and 25\%, while for PM-64QAM, a FEC OH between 32\% and 37\% is necessary to obtain similar data rates. At 6000~km, PM-64QAM can achieve approximately 334 Gbits/symbol with a respective FEC OH of 44\%. For the same net rate, the optimized constellation has a reach increase of approximately 520 km (8.7\%) with respect to PM-64QAM and 340 km (5.5\%) with respect to PM-PS64QAM, while demanding a FEC OH of 33\%. The increased distance indicates a better nonlinear tolerance for the optimized constellation, while also keeping a lower FEC complexity than PM-64QAM.

Other 4D modulation formats can be found on the literature, for example in \cite{ErikSP}. However, as reported in \cite{Alex2015}, these constellations perform suboptimally in terms of GMI when compared to PM-$M$QAM formats with the same number of bits/symbol. In addition, finding a binary labeling for those 4D constellations is challenging \cite{Alex2015}.


\section{Conclusions}\label{sc:conc}

We have presented a 4D 10 bits/symbol modulation format which outperforms standard PM-32QAM and PM-64QAM at distances greater than 3500 km. The proposed constellation was obtained via an autoencoder structure, where the split-step Fourier method was used as the channel model. The gains were mainly driven by geometric shaping, since the symbol probabilities converged to a uniform distribution. The resulting symbols had a wide energy variation, contradicting forced constant energy approaches used in the literature. Future works include developing 4D modulation formats with even more than 10 bits/symbol and to include additional fiber impairments.

\section*{Acknowledgments}
This work is supported by the Netherlands Organisation for Scientific Research (NWO) via the VIDI Grant ICONIC (project number 15685). The work of A. Alvarado has received funding from the European Research Council (ERC) under the European Union's Horizon 2020 research and innovation programme (grant agreement No 757791). The work of G.~Liga is supported by the EuroTechPostdoc programme under the European Union's Horizon 2020 research and innovation programme (grant agreement No 754462).

\ifCLASSOPTIONcaptionsoff
  \newpage
\fi

\balance

\bibliographystyle{IEEEtran}
\bibliography{IEEEabrv,Bibliography}

\vfill

\end{document}